**Title Page**

# A distribution-free change-point monitoring scheme in high-dimensional settings with application to industrial image surveillance

**(Preprint version)**

**Author 1: *Corresponding author**

*Niladri Chakraborty

Email: niladri.chakraborty30@gmail.com

**Affiliation:**

Department of Mathematical Statistics and Actuarial Science, University of the Free State, 205 Nelson Mandela Dr, Park West, Bloemfontein, South Africa 9301.

ORCID ID: 0000-0001-9853-4067

**Author 2:**

Chun Fai Lui

Email: chunflui5-c@my.cityu.edu.hk

**Affiliation:**

Department of Advanced Design and Systems Engineering, City University of Hong Kong, 83 Tat Chee Ave, Kowloon Tong, Hong Kong, Hong Kong.

ORCID ID: 0000-0002-2379-0821

**Author 3:**

Ahmed Maged

Email: amaged2-c@my.cityu.edu.hk

**Affiliation:**

Department of Advanced Design and Systems Engineering, City University of Hong Kong, 83 Tat Chee Ave, Kowloon Tong, Hong Kong, Hong Kong.

ORCID ID: 0000-0002-5071-5253

# A distribution-free change-point monitoring scheme in high-dimensional settings with application to industrial image surveillance

(Preprint version)

**Abstract:** Existing monitoring tools for multivariate data are often asymptotically distribution-free, computationally intensive, or require a large stretch of stable data. Many of these methods are not applicable to 'high-dimension, low sample size' scenarios. With rapid technological advancement, high-dimensional data has become omnipresent in industrial applications. We propose a distribution-free change-point monitoring method applicable to high-dimensional data. Through an extensive simulation study, performance comparison has been done for different parameter values, under different multivariate distributions with complex dependence structures. The proposed method is robust and efficient in detecting change-points under a wide range of shifts in the process distribution. A real-life application illustrated with the help of high-dimensional image surveillance dataset.

**Keywords**: High-dimensional data; Distribution-free monitoring; Change-point; Image monitoring; Run length.

## 1. Introduction

In todays' world, advancement in data acquisition technologies have made data-rich applications quintessential. In such applications, it is important to monitor and detect anomalies in high-dimensional settings; for example, anomalous event detection in networks (Wang and Xie 2021) , quality monitoring in additive manufacturing (Najjartabar Bisheh et al. 2021), monitoring spatio-temporal data arising from solar flare activity (Yan et al. 2018). It is well known that distributional assumptions for a sequence of observations are restrictive and yet more difficult to justify in case of multivariate data (Chen et al. 2016; Mukherjee and Marozzi 2022). For example, multivariate normal assumption in traditional Hotelling $T^2$ chart may fail in monitoring multivariate data with complex dependence structure. Even when the asymptotic normality is ensured, it only works around the central part of the process distribution. In contrast, we are more concerned about the tail behavior of a process distribution in process monitoring applications. Moreover, the traditional Hotelling $T^2$ chart is not applicable when the data dimension $p$ is close to the sample size $n$, or even higher, due to singularity of covariance matrix.

Literature on distribution-free high-dimensional process monitoring is sparse. Over the last decade, a number of authors have made attempts to circumvent restrictive assumptions in parametric multivariate monitoring methods. For example, Chen et al. (2016), Shu & Fan (2018), Li et al. (2020), Kurt et al. (2021), Chen and Wang (2022). A recent account on multivariate process monitoring can be found in Sofikitou & Koutras (2020). Literature review suggests that the existing methods to evade the parametric assumptions are often, (i) computationally intensive, (ii) asymptotically distribution-free, or (iii) rely upon the availability of a large stretch of training data without any distributional change.

In high-dimensional setting, process distribution parameters are often unknown, and estimation of parameters is an intricate practice. For instance, estimation of covariance matrix in high-dimensional setting is itself an important topic of research on its own merits. In practice, substantial amount of high-dimensional training data may not be easy to obtain. Consequently, a distribution-free, high-dimensional, monitoring method that does not require known process parameters or a large training data would be certainly desirable. In this article, we propose a distribution-free change-point method to monitor high-dimensional data. This is an original manuscript of an article

We emphasize that the proposed method is, i. distribution-free; ii. effective in high-dimensional settings; iii. affine invariant; iv. computationally efficient; v. exempted from the need of a large training data. Unlike existing high-dimensional monitoring methods, the proposed method offers all five important benefits together.

The rest of the article is organized as follows. Section 2 is devoted to the statistical framework of the proposed method. In Section 3, in-control design and robustness of the proposed method are discussed. Section 4 confers the study on out-of-control performance of the proposed method. An illustrative example of the monitoring of high-dimensional image data is provided in Section 5. Finally, some concluding remarks are provided in Section 6.

## 2. Statistical framework

A real valued $p$-dimensional sequence of random vectors $\{\boldsymbol{Y_1}, \boldsymbol{Y_2}, \boldsymbol{Y_3}, \dots, \boldsymbol{Y_n}, \dots\}$ is obtained at regular time intervals from an online production process/system. As a hypothesis testing problem, the sequential monitoring of change-point can be written as,

$$\boldsymbol{H_0}: \boldsymbol{Y_i} \sim F^p \text{ for } i = 1, 2, 3, \dots \quad \textbf{vs.}$$

$$\boldsymbol{H_1}: \boldsymbol{Y_i} \sim F^p \text{ for } i < \tau, \text{ and } \boldsymbol{Y}_i \sim G^p \text{ for } i \geq \tau. \qquad (1)$$

where $F^p = F^p(\boldsymbol{\mu_0})$ and $G^p = G^p(\boldsymbol{\mu_1})$ are two multivariate distributions with $\boldsymbol{\mu_0}$ and $\boldsymbol{\mu_1}$ being $p$-dimensional parameters before and after the change-point $\tau$, respectively. In real-life applications, we may have no prior knowledge about $F^p$ or $G^p$. We say that there is no change-point when $F^p = G^p$. A shift in the process may occur as $\boldsymbol{\mu_0} \neq \boldsymbol{\mu_1}$, or a more general change in the distributional form $F^p \neq G^p$ may take place.

In this article, we consider a $\mathcal{L}_2$-norm based change-point monitoring scheme. In retrospective testing and online monitoring of high-dimensional data, $\mathcal{L}_2$-norm transformation is well accepted in literature (Jurečková and Kalina 2012; Mukherjee and Marozzi 2022). Marozzi (2015) compared several distance measures for high-dimensional two-sample location problem and concluded that the $\mathcal{L}_2$-norm performs close, and is often superior to other distance measures. Modarres & Song (2020) underlined the interpoint $\mathcal{L}_2$-norm measures as the keystone in several multivariate techniques such as clustering, classification, and comparison of distributions. Motivated by these developments, we discuss change-point monitoring scheme based on the $\mathcal{L}_2$-norm transformation. Other distance measures are not considered in this article.

In order to monitor the sequence $\{\boldsymbol{Y_1}, \boldsymbol{Y_2}, \boldsymbol{Y_3}, \dots, \boldsymbol{Y_n}, \dots\}$**,** we consider a *moving window* approach based on $\mathcal{L}_2$-norm of the random vectors $\boldsymbol{Y}_i$. Next, we describe a general monitoring algorithm, and then we discuss the test statistic used in the process.

**Algorithm for the $\boldsymbol{\mathcal{L}_2}$-norm based method:**

i. For every $\boldsymbol{Y}_i$, $\mathcal{L}_2$-norm is obtained as $d_i = ||\boldsymbol{Y_i}||$, for $i = 1, 2, 3, \dots$ ;

ii. For the sequence of $\mathcal{L}_2$-norm values $\{d_1, d_2, \dots, d_n, \dots\}$, let us consider a subset (window) of size $w$ given by $S_i = \{d_i, d_{i+1}, \dots, d_{i+w-1}\}$, $i = 1, 2, 3,\dots$;

iii. $S_i$ is partitioned into two non-overlapping subsamples $S_{il}^1 = \{d_i, \dots, d_{i+l-1}\}$ and $S_{il}^2 = \{d_{i+l}, \dots, d_{i+w-1}\}$, for $l_0 \leq l \leq (w - l_0)$, $l_0$ is the first partition point of the $i^{th}$ window, called the *quarantine constant* (QC). Note that there are $(w - 2l_0 + 1)$ partitions for every window of size $w$;

iv. A two-sample nonparametric one-sided statistic $T_{il}$ is obtained for the $l^{th}$ partition of the $i^{th}$ window, $l = 1, 2, \ldots, (w - 2l_0 + 1)$. Details about $T_{il}$ are discussed later;

v. Then we calculate a statistic $T_i$ as a function of $\{T_{il}; l = 1, 2, \ldots, (w - 2l_0 + 1)\}$;

vi. Steps (ii) – (v) are to be repeated for every time instance $i = 1, 2, 3, \ldots$ until $T_i \geq h_i$ for some appropriate threshold $h_i$.

**Distance-based statistic $T_{il}$:**

For the $l^{th}$ partition of the $i^{th}$ window, $S_{il}^1 = \{d_i, \ldots, d_{i+l-1}\}$ and $S_{il}^2 = \{d_{i+l}, \ldots, d_{i+w-1}\}$, let us denote the distance values $d_j^1 \in S_{il}^1$, $j = 1, 2, \ldots, l$, and $d_j^2 \in S_{il}^2$, $j = 1, 2, \ldots, (w - l)$. Ng & Balakrishnan (2005) proposed a weighted-precedence statistic that is computationally simple and overall superior to several other precedence-type tests. Precedence-type tests are useful when process distribution or process parameters are unknown (see Ng and Balakrishnan (2005) and the references therein). We obtain the frequencies of $d_j^1 \in S_{il}^1$ between two consecutive $d_{(k)}^2$, i.e., the ordered $d_k^2 \in S_{il}^2$ values. Then we take the weighted sum of the frequencies and divide that by $(l(w - l))$. A modified weighted-precedence statistic, thus, can be obtained as,

$$T_{il} = \frac{\sum_{k=1}^{(w-l+1)} (w-l-k+1) \sum_{j=1}^{l} I\left(d_{(k-1)}^2 < d_j^1 \leq d_{(k)}^2\right)}{l(w-l)}, \quad (2)$$

where $d_{(0)}^2 = 0$ and $d_{(w-l+1)}^2 = \infty$, and $I(.)$ is an indicator function so that $I(C) = 1$ if the condition $C$ is true, and 0, otherwise. Note that $0 \leq T_{il} \leq 1$ with $T_{il} = 1$ when $d_j^1 \leq d_{(1)}^2$ for all $j$, and $T_{il} = 0$ when $d_j^1 > d_{(w-l)}^2$ for all $j$. For any extreme $T_{il}$, large or small, there may be a significant difference in distributions of $S_{il}^1$ and $S_{il}^2$. Let us consider $T_{i(0.25)}$ = 1st quartile of $\{T_{il}; l_0 \leq l \leq (w - l_0)\}$ and $T_{i(0.75)}$ = 3rd quartile of $\{T_{il}; l_0 \leq l \leq (w - l_0)\}$ for the $i^{th}$ window. Then we define a statistic $T_i$ as

$$T_i = \max\{T_{i(0.75)}, 1 - T_{i(0.25)}\}. \quad (3)$$

We consider a one-sided statistic as it seems easy to implement and interpret in the context of change-point detection. A control limit for $T_i$ is needed to determine whether the process is in-control (IC) with no change-point $\tau$, or out-of-control (OOC) with a change-point $\tau$ in the data sequence. Since $T_i$ is defined on a distribution-free statistic, the proposed method is also distribution-free. Note that, the sequence of random vectors $\{\boldsymbol{Y}_1, \boldsymbol{Y}_2, \boldsymbol{Y}_3, \dots, \boldsymbol{Y}_n, \dots\}$ are taken to be temporally independent, but the test statistic values $\{T_1, T_2, \dots, T_n, \dots\}$ are not temporally independent as every pair of windows $S_i$, for all $i$, are overlapping. Because of this temporal dependency, the control limits would be different for different windows. Thus, we would get a sequence of control limits $\{h_1, h_2, \dots, h_n, \dots\}$.

To construct the plotting statistic $T_i$, we consider $T_{i(0.75)}$ and $T_{i(0.25)}$ to avoid any possible outlier effect caused by the extreme order statistics such as maximum or minimum. But, when we get a signal at a certain window, we consider the partition containing maximum or minimum of $T_{il}$ from that window to estimate the change-point. The reason being that when we get a signal, we can sense a possible change-point lying somewhere in that window. The partition point with the maximum/minimum of $T_{il}$ may be taken as a good estimate of the change-point. Note that, the proposed method is applicable to high-dimensional data as it does not require the number of sample to be larger than the dimension $p$. When there is a signal at the $r^{th}$ window, the estimated change-point is obtained as

$$\hat{\tau} = r + l_0 + l_{cp} - 1, \quad (4)$$

where $l_{cp}$ is the partition point with the maximum or minimum of $\{T_{il}; l_0 \leq l \leq (w - l_0)\}$.

## 3. In-control design and robustness

In order to implement the proposed method, it is important to obtain the control limits so that there is a small probability of a false signal when there is no change-point in the data sequence. It is also important to investigate the IC robustness of the proposed method.

### 3.1. In-control design

Run length ($RL$) is a popular measure in literature to design and evaluate the operating characteristics of monitoring procedures (Holland and Hawkins 2014; Mukherjee and Marozzi 2022). It is conventional to use the median and average $RL$, denoted by $MRL$ and $ARL$, respectively, for design and performance study. In the present context, $ARL$ and $MRL$ would provide the average and median number of windows $R = r$ required to get $T_r \geq h_r$ given $T_j < h_j$ for all $j = 1, 2, 3, \ldots, (r-1)$. For brevity, other performance measures are not considered in this article. Note that a control limit for $T_i$, for any $i$, would not work for $T_{i+1}$ because of the temporal dependency. Therefore, a sequence of control limits $\{h_1, h_2, \ldots, h_n, \ldots\}$ is obtained so that,

$$P[T_1 \geq h_1] = \alpha,$$

$$P\left[T_i \geq h_i \mid T_1 < h_1, \ldots, T_{(i-1)} < h_{(i-1)}\right] = \alpha, \text{ for } i = 2, 3, 4, 5, \ldots \quad (5)$$

where $\alpha = \frac{1}{ARL_0}$. In many process monitoring applications, the accepted IC average $RL$ is considered as $ARL_0 \approx 200$ (Xiang et al. (2020), Yan et al. (2018)). We choose $\alpha \approx 0.004$ so that the IC $MRL$, denoted by $MRL_0 \approx 170$ and the corresponding $ARL$, denoted by $ARL_0 \approx 245$. Other values of $MRL_0$ and $ARL_0$ can be considered as well to obtain the control limits following the algorithm discussed in Section 2.

An analytical solution for (5) is mathematically intractable. Therefore, we perform Monte-Carlo simulation to obtain the control limits. In each iteration, 2500 random samples $\{\boldsymbol{Y}_1, \boldsymbol{Y}_2, \boldsymbol{Y}_3, \dots, \boldsymbol{Y}_{2500}\}$ are drawn from a $p$-dimensional normal distribution with endpoints as standard normal distributions. Because the proposed method is distribution-free, the control limits are not going to be affected by the underlying process distribution. Then, the sequence of $\mathcal{L}_2$-norm values $\{d_1, d_2, \dots, d_n, \dots\}$ are obtained for each sample of size 2500. After that, for a specific $w$ and $l_0$, the test statistic $T_i$ values are obtained for every sequence $\{d_1, d_2, \dots, d_n, \dots\}$. This is repeated for 10000 times, and then using these 10000 sequences of $\{T_i\}$, control limits $\{h_i\}$ are obtained so that the conditions in Equation (5) are satisfied. For each of these 10000 simulation runs, run length $R_k = r_k$ is obtained, for $k = 1, 2, \dots, 10000$, so that there is a signal at $r_k$ at the $k^{th}$ simulation run. Finally, $MRL_0$ and $ARL_0$ are obtained from the set of $RL$ observations $\{r_1, r_2, \dots, r_{10000}\}$. For the proposed method, computation of 5000 plotting statistics for a sequence of multivariate normal data of dimension 100 takes approximately 3 seconds with a standard computer with core i7 processor. Therefore, the proposed method is not computationally expensive, even for a very high-dimensional data.

In Table 1, control limits $h_i$ are displayed for window size $w = 15, 17, 20$, and quarantine constant $l_0$ = 2, 3, 4, 5, for multivariate normal distribution. For brevity, we only present a few $h_i$ values in Table 1. Other values of control limits are available from the authors on request. The corresponding $MRL_0$ and $ARL_0$ values are presented in Table 2. The mean vector of the multivariate normal distribution is taken as a vector of 0's of dimension $p$. The covariance matrix $\boldsymbol{C}$ is considered as follows: the $(i,j)^{th}$ element of the matrix $\boldsymbol{C}$ is given by

$$C_{ij} = \frac{\min(i,j)}{\max(i,j)} \sigma_i \sigma_j, \text{ for } i, j = 1, 2, \dots, p, \tag{6}$$

where $\sigma_i = c_0 + (i-1)/(p-1)$ and the constant $c_0$ is taken as 0.5.

**Table 1. Control limits, $h_i$, for multivariate normal distribution, for $w = 15, 17, 20$, and $l_0 = 2, 3, 4, 5$.**

| | **$w = 15$** | | | | **$w = 17$** | | | | **$w = 20$** | | | |
|---|---|---|---|---|---|---|---|---|---|---|---|---|
| | **$l_0$** | | | | **$l_0$** | | | | **$l_0$** | | | |
| **$i$** | **2** | **3** | **4** | **5** | **2** | **3** | **4** | **5** | **2** | **3** | **4** | **5** |
| 1 | 0.9306 | 0.9259 | 0.9266 | 0.9182 | 0.9043 | 0.9036 | 0.8958 | 0.9023 | 0.8889 | 0.8780 | 0.8700 | 0.8679 |
| 2 | 0.9156 | 0.9116 | 0.9067 | 0.9132 | 0.9001 | 0.8902 | 0.8816 | 0.8847 | 0.8802 | 0.8669 | 0.8646 | 0.8544 |
| 3 | 0.9185 | 0.9122 | 0.9121 | 0.9076 | 0.8946 | 0.8815 | 0.8758 | 0.8691 | 0.869 | 0.8633 | 0.8533 | 0.8524 |
| 4 | 0.9180 | 0.904 | 0.9023 | 0.8976 | 0.896 | 0.8819 | 0.8777 | 0.8747 | 0.8637 | 0.8646 | 0.8533 | 0.8489 |
| 5 | 0.9121 | 0.9114 | 0.8971 | 0.8996 | 0.895 | 0.8811 | 0.8753 | 0.8842 | 0.8622 | 0.8598 | 0.8533 | 0.849 |
| 10 | 0.9177 | 0.9071 | 0.8963 | 0.8954 | 0.8917 | 0.8852 | 0.8803 | 0.8754 | 0.8627 | 0.8518 | 0.8571 | 0.8454 |
| 20 | 0.9098 | 0.903 | 0.8936 | 0.8889 | 0.8954 | 0.8793 | 0.8681 | 0.8704 | 0.8627 | 0.8527 | 0.8438 | 0.8445 |
| 30 | 0.9208 | 0.9053 | 0.8974 | 0.906 | 0.8913 | 0.8833 | 0.8718 | 0.8727 | 0.8627 | 0.8545 | 0.8462 | 0.8416 |
| 40 | 0.9122 | 0.904 | 0.9019 | 0.8956 | 0.8917 | 0.8764 | 0.8734 | 0.8801 | 0.863 | 0.8615 | 0.8462 | 0.8395 |
| 50 | 0.9119 | 0.9122 | 0.8996 | 0.892 | 0.8856 | 0.8805 | 0.8757 | 0.8744 | 0.8611 | 0.8529 | 0.8462 | 0.8431 |
| 100 | 0.9113 | 0.904 | 0.9023 | 0.8954 | 0.8959 | 0.8783 | 0.8746 | 0.8689 | 0.8669 | 0.8497 | 0.8462 | 0.8502 |
| 500 | 0.9126 | 0.9043 | 0.8949 | 0.8972 | 0.8958 | 0.8857 | 0.8738 | 0.8697 | 0.8687 | 0.8579 | 0.8485 | 0.8445 |
| 1000 | 0.9180 | 0.8969 | 0.905 | 0.8998 | 0.8998 | 0.8882 | 0.8776 | 0.8697 | 0.8646 | 0.8498 | 0.8452 | 0.849 |
| 2000 | 0.9209 | 0.9097 | 0.895 | 0.8998 | 0.8901 | 0.8834 | 0.8811 | 0.8754 | 0.8611 | 0.8579 | 0.8533 | 0.8431 |

**Table 2. $MRL_0$ and $ARL_0$ values for multivariate normal distribution, for window size $w$ = 10, 13, 15, 17, 20, and the quarantine constant $l_0$ = 2, 3, 4, 5.**

| $p$ | $l_0$ | $ARL_0$ $w$ =10 | $w$ =13 | $w$ =15 | $w$ =17 | $w$ =20 |
|---|---|---|---|---|---|---|
| 10 | 2 | 191.69 | 240.01 | 245.95 | 241.15 | 237.36 |
| | 3 | 182.70 | 240.14 | 245.23 | 243.99 | 247.78 |
| | 4 | 194.17 | 240.42 | 247.14 | 245.13 | 234.68 |
| | 5 | 151.79 | 233.67 | 244.43 | 243.71 | 253.16 |
| 25 | 2 | 198.39 | 231.89 | 244.61 | 243.04 | 235.31 |
| | 3 | 185.01 | 236.12 | 244.12 | 243.14 | 251.19 |
| | 4 | 191.74 | 238.55 | 243.44 | 241.61 | 234.71 |
| | 5 | 153.02 | 233.74 | 244.67 | 246.89 | 250.08 |
| 50 | 2 | 195.46 | 233.70 | 245.47 | 239.82 | 231.84 |
| | 3 | 183.05 | 236.89 | 242.95 | 236.48 | 245.02 |
| | 4 | 191.72 | 237.75 | 243.77 | 241.45 | 232.59 |
| | 5 | 152.30 | 234.46 | 246.79 | 244.29 | 244.44 |
| 100 | 2 | 195.38 | 235.76 | 246.46 | 243.92 | 237.24 |
| | 3 | 181.14 | 242.52 | 245.76 | 236.99 | 243.99 |
| | 4 | 195.34 | 238.83 | 241.92 | 248.07 | 234.76 |
| | 5 | 147.70 | 228.79 | 249.07 | 245.86 | 248.67 |

| $p$ | $l_0$ | $MRL_0$ $w$ =10 | $w$ =13 | $w$ =15 | $w$ =17 | $w$ =20 |
|---|---|---|---|---|---|---|
| 10 | 2 | 135 | 167 | 170 | 167 | 166 |
| | 3 | 127 | 165 | 171 | 170 | 172 |
| | 4 | 133 | 169 | 171 | 171 | 162 |
| | 5 | 108 | 161 | 171 | 168 | 171 |
| 25 | 2 | 138 | 161 | 169 | 175 | 163 |
| | 3 | 129 | 166 | 169 | 172 | 177 |
| | 4 | 134 | 163 | 170 | 169 | 160 |
| | 5 | 107 | 158 | 170 | 171 | 176 |
| 50 | 2 | 136 | 160 | 172 | 172 | 162 |
| | 3 | 129 | 165 | 170 | 161 | 173 |
| | 4 | 134 | 165 | 167 | 166 | 160 |
| | 5 | 105 | 158.5 | 169 | 172 | 170 |
| 100 | 2 | 135 | 161 | 173 | 172 | 163 |
| | 3 | 126 | 170.5 | 173 | 162 | 170 |
| | 4 | 137 | 166 | 167.5 | 165 | 163 |
| | 5 | 103 | 159 | 171 | 172 | 172 |

### 3.2. Robustness study

To assess the robustness of the proposed method, control limits obtained for a multivariate normal distribution with $p$ = 25 are used for other distributions to obtain $MRL_0$ and $ARL_0$ as reported in Table 3. We consider the window size $w = 15$, and the quarantine constant $l_0$ = 3. For standard symmetric multivariate distributions, we consider multivariate normal distribution, $t_5$-distribution, and the Cauchy distribution. For multivariate $t$ and the Cauchy distribution, the scale matrix is considered as the same as the covariance matrix $\boldsymbol{C}$ in Equation (6). We consider a multivariate skewed distribution with end points as exponential distributions with rate 1, without loss of generality, connected by Gaussian copula. An R package called 'copula' is used to draw random samples from the multivariate exponential distribution. Copulas are popular tools to model complex dependencies among the endpoints of multivariate distributions. For more details on copula modelling, one may see Trivedi & Zimmer (2005).

For the same set of control limits obtained for $p = 25$, $w = 15$ and $l_0 = 3$, the estimated $ARL_0$ and $MRL_0$ are close to the nominal value for all distributions under consideration. The values in Table 3 display the IC robustness of the proposed method in small sample setting. Holland & Hawkins (2014) concluded that a monitoring method is IC robust when the estimated $ARL_0$ falls within 10% of the nominal value. We can conclude that the proposed method attains a better IC robustness compared to Holland & Hawkins (2014), with estimated $ARL_0$ and $MRL_0$ falling within 2% of the nominal value. Clearly, the proposed method is not influenced by the process density and dimensions.

**Table 3.** $\boldsymbol{MRL_0}$ **and** $\boldsymbol{ARL_0}$ **values for** $\boldsymbol{p} = $ **25, 50, 100, 20000,** $(\boldsymbol{l_0}, \boldsymbol{w})$ **= (3, 15), (2, 17) under different multivariate distributions.**

| Density | $p$ | $MRL_0$ | $ARL_0$ | $MRL_0$ | $ARL_0$ |
|---|---|---|---|---|---|
| | | $l_0 = 3, w = 15$ | | $l_0 = 2, w = 17$ | |
| **Normal** | 25 | 172.00 | 246.93 | 175.00 | 243.04 |
| $t_5$ | 25 | 169.00 | 244.64 | 168.00 | 247.69 |
| **Cauchy** | 25 | 170.00 | 245.90 | 171.00 | 242.50 |
| **Exponential** | 25 | 170.00 | 244.98 | 176.00 | 247.32 |
| **Normal** | 50 | 169.00 | 243.34 | 172.00 | 239.82 |
| $t_5$ | 50 | 169.00 | 246.88 | 158.00 | 235.60 |
| **Cauchy** | 50 | 172.00 | 247.89 | 172.00 | 242.62 |
| **Exponential** | 50 | 170.00 | 239.05 | 174.49 | 246.97 |
| **Normal** | 100 | 174.57 | 246.28 | 172.41 | 243.92 |
| $t_5$ | 100 | 174.00 | 248.99 | 176.00 | 248.59 |
| **Cauchy** | 100 | 172.00 | 245.43 | 168.00 | 246.77 |
| **Exponential** | 100 | 174.00 | 246.73 | 180.00 | 252.50 |
| **Normal** | 20000 | 171.00 | 244.73 | 168.00 | 245.98 |
| $t_5$ | 20000 | 172.21 | 247.17 | 175.00 | 242.99 |
| **Cauchy** | 20000 | 168.00 | 245.21 | 172.00 | 242.15 |
| **Exponential** | 20000 | 171.00 | 243.68 | 175.00 | 243.87 |

### 3.3. Remarks about affine invariance

Let us consider the $l^{th}$ partition of the $i^{th}$ window, $S_{il}^1 = \{d_i, \ldots, d_{i+l-1}\}$ and $S_{il}^2 = \{d_{i+l}, \ldots, d_{i+w-1}\}$, and denote the distance values $d_j^1 \in S_{il}^1$, $j = 1, 2, \ldots, l$, and $d_j^2 \in S_{il}^2$, $j = 1, 2, \ldots, (w - l)$. For any $\boldsymbol{Y}_i$, we define the affine transformed data as $\boldsymbol{Y}_j^* = \mathcal{T}(\boldsymbol{Y}_j) = \boldsymbol{A}\boldsymbol{Y}_j + \boldsymbol{b}$, where $\mathcal{T}(.)$ is an affine transformation with $\boldsymbol{A}$ being a nonsingular matrix and $\boldsymbol{b}$ being a $p$-dimensional vector. Let us also denote $d_i^* = ||\boldsymbol{\mathcal{T}}(\boldsymbol{Y_i})||$, for $i = 1, 2, 3, \ldots$ $\mathcal{L}_2$-norm transformation blended with the precedence test implemented in a change-point setting provides affine invariance for the proposed method if the same transformation is applied on the two datasets. A short proof is provided in the Appendix. Weighted-precedence statistic is a modification of the classical precedence statistic (Ng and Balakrishnan 2005). Note that, we do the same affine transformation on both $S_{il}^1$ and $S_{il}^2$. Therefore, from (2), the modified weighted-precedence statistics $T_{il}$ for $\{d_i\}$ and $T_{il}^*$ for $\{d_i^*\}$ would be same even after taking an affine transformation of the dataset. This implies that the proposed method is affine invariant.

## 4. Performance evaluation

We carry out a performance study under different OOC scenarios based on $MRL_1$ for a number of multivariate symmetric and skewed distributions. $MRL_1$ is less affected by outliers and therefore, provide more stable results than $ARL_1$.

### 4.1. Performances under various change-point scenario

To inspect early change detection ability, we assume the true change point $\tau = 10$ in the data sequence. A multivariate exponential distribution of dimension $p = 20$ with rate parameter $\lambda_1 =$ 0.0001, 0.01, 3, 5, is considered as a shifted process distribution after $\tau = 10$. The end points are assumed to be connected by Gaussian copula and Clayton copula (Trivedi & Zimmer, 2005). The IC Clayton copula parameter is taken as $\xi = 1$. The window size is $w = 15$, and the quarantine constant $l_0$ = 3, 4, 5. When a signal is raised, we obtain the partition $l_{cp}$ containing the extremum value of the chosen statistic and estimate the change-point as in Equation (4) for every simulation run. The median of the estimated the change-points is taken as the estimated change-point $\hat{\tau}$.

In Table 4, the $MRL_1$ values are presented for changes in $\lambda$ and the dependence structure of multivariate exponential, for $p = 20$, $w = 15$, at true change points $\tau = 10$, 50. Note that, the proposed method has close performance in change-point detection for $l_0 = 3$, 4, 5. Especially, for Gaussian copula, it shows a better performance compared to Clayton copula. In Fig. 1, the OOC performance is displayed in more detail for the multivariate exponential connected by Clayton copula. In Fig. 1, the quarantine constant (QC) is taken as $l_0 = 3$, 5, and the exponential rate is varied over $\lambda_1 = 0.0001$, 0.005, 0.01, 0.05, 0.1, 0.2, 0.3, 0.5, 1, 1.5, 2, 2.5, 3, 3.5, 4, 4.5, 5.

**Table 4. $MRL_1$ values for changes in $\lambda$ and the dependence structure of multivariate exponential with Gaussian and Clayton copula, for $p = 20$, $w = 15$, at true change point $\tau =$ 10, 50.**

**Changes in $\lambda$**

| | Gaussian copula | | | | | | Clayton copula | | | | | |
|---|---|---|---|---|---|---|---|---|---|---|---|---|
| | | | | | | $\tau$ =10 | | | | | | |
| $\lambda_1$ | $l_0 = 3$ | $\hat{\tau}$ | $l_0$= 4 | $\hat{\tau}$ | $l_0 = 5$ | $\hat{\tau}$ | $l_0 = 3$ | $\hat{\tau}$ | $l_0$= 4 | $\hat{\tau}$ | $l_0 = 5$ | $\hat{\tau}$ |
| 0.0001 | 1 | 11 | 1 | 11 | 2 | 11 | 1 | 11 | 2 | 11 | 2 | 11 |
| 0.01 | 1 | 11 | 1 | 11 | 2 | 11 | 1 | 11 | 2 | 11 | 2 | 11 |
| 3 | 1 | 11 | 1 | 11 | 2 | 11 | 79 | 86 | 59 | 66 | 57 | 64.5 |
| 5 | 1 | 11 | 1 | 11 | 2 | 11 | 6 | 13 | 5 | 12 | 4 | 11 |
| | | | | | | **Changes in $\lambda$ and copula structure (Gaussian to Clayton)** | | | | | | |
| | | | $\tau$ =10, $\xi = 2$ | | | | | | $\tau$ =10, $\xi = 5$ | | | |
| 0.001 | 1 | 11 | 1 | 11 | 1 | 11 | 1 | 11 | 1 | 11 | 1 | 11 |
| 5 | 1 | 11 | 1 | 11 | 1 | 11 | 1 | 11 | 1 | 11 | 1 | 11 |
| | | | $\tau$ =50, $\xi = 2$ | | | | | | $\tau$ =50, $\xi = 5$ | | | |
| 0.001 | 42 | 51 | 41 | 51 | 41 | 51 | 40 | 51 | 40 | 50 | 41 | 51 |
| 5 | 40 | 51 | 40 | 51 | 41 | 51 | 40 | 51 | 41 | 51 | 41 | 51 |

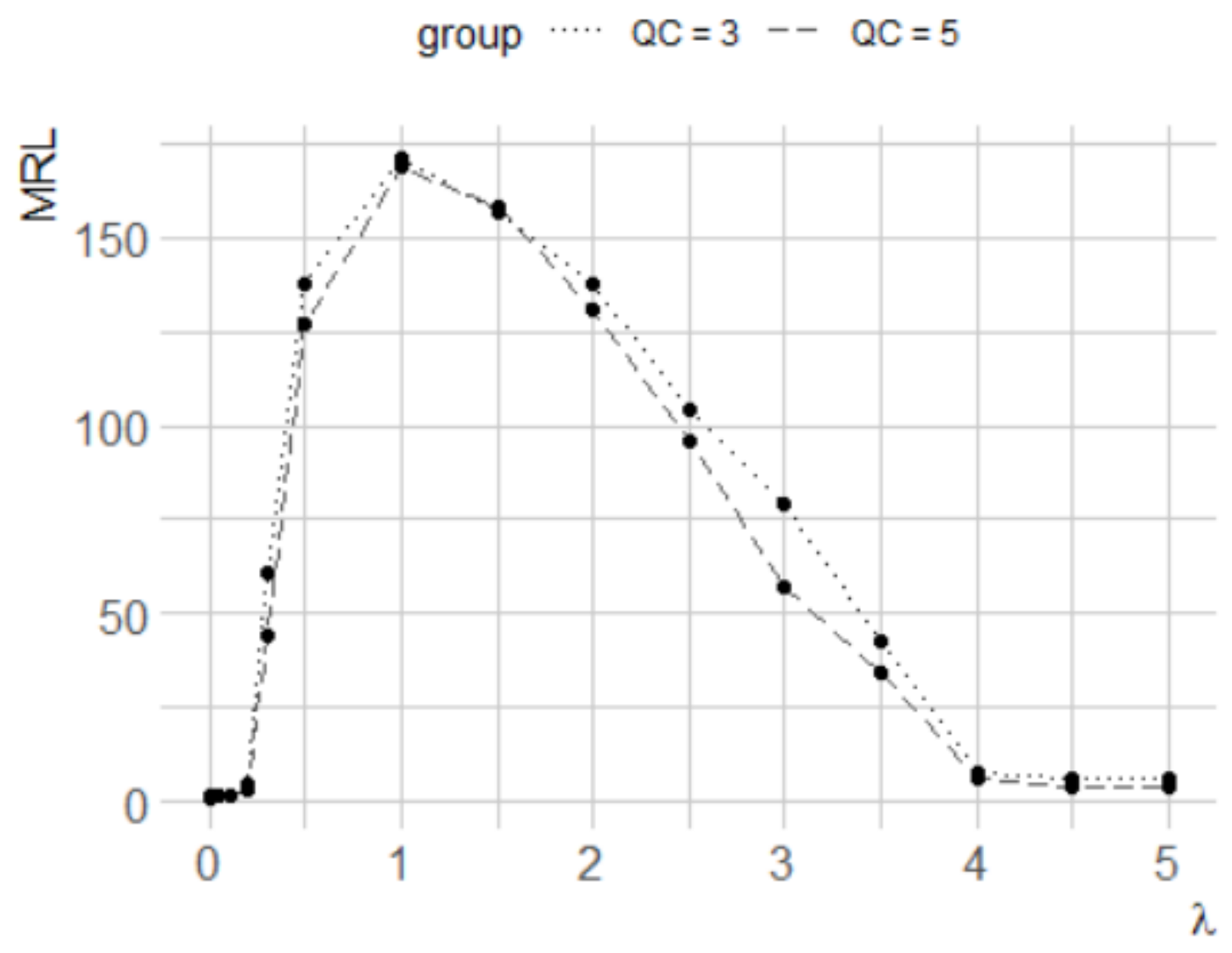

**Fig. 1.** $MRL_1$ **values for multivariate exponential connected with Clayton copula for different** $\lambda_1$ **and quarantine constant (QC)** $l_0$**.**

In Table 4, we also present $MRL_1$ values when simultaneous changes took place in the dependence structure and the marginal exponential rate. The dependence structure is assumed to change from Gaussian to Clayton copula with parameter $\xi$ = 2, 5. A larger $\xi$ implyies a stronger dependence among the end points (see the Appendix). It can be noted that the proposed method is able to detect simultaneous change in the exponential rate and the dependence structure. The estimated change-points in Tables 4 are remarkably close to the true change-points for a wide range of shifts.

Next, we examine the OOC performance for shifted multivariate normal distributions with $p =$ 50, 100, in Table 5. For the mean shift, we consider $\delta_\mu = 1.5$ in each coordinate of the mean vector. For the OOC normal distribution with covariance shift, we generate a positive definite covariance matrix using the R package 'clusterGeneration' with variance range (0.1, 0.5) and $\alpha_d = 1.5$ (regarded as Case I), and variance range (1.5, 2.5) and $\alpha_d$ = 1.5 (regarded as Case II). The IC covariance is taken the same as the covariance matrix described in Equation (6), before the change-point $\tau$ = 25, 50, 100. For the simultaneous changes in mean and covariance, we consider $\delta_\mu =$ 0.5, 1.5, and covariance matrices as Case I and Case II. It can be observed that the proposed method is efficiently able to raise a signal for a change in the mean and covariance, except for the case with mean shift 0.5 and 'Case I' shift in covariance matrix. In fact, for a change in the dependence structure, relative order of the observations from two subsamples (Step (iii) of the algorithm in Section 2) is disturbed.

**Table 5: $MRL_1$ values for multivariate normal with shift in the mean and the covariance matrix for $w = 15$, $l_0 = 3$.**

| | $\tau = 25$ | $\tau = 50$ | $\tau = 100$ |
|---|---|---|---|
| **Mean shift** | | $p = 50$ | |
| 0.7 | 163 | 162 | 160 |
| 0.9 | 142 | 147 | 146 |
| 1.1 | 113 | 108.5 | 116 |
| 1.3 | 77 | 74 | 94 |
| 1.5 | 21 | 45 | 92 |
| 1.7 | 18 | 43 | 91 |
| 2.0 | 16 | 41 | 91 |
| | **Cov. matrix shift** | | |
| Case 1 | 20 | 44 | 92 |
| Case 2 | 21 | 45 | 92 |
| | **Mean and Cov. Matrix shift** | | |
| 0.5 and Case 1 | 148 | 144 | 147 |
| 0.5 and Case 2 | 19 | 43 | 91 |
| 1.5 and Case 1 | 17 | 41 | 91 |
| 1.5 and Case 2 | 16 | 41 | 90 |
| **Mean shift** | | $p = 100$ | |
| 0.7 | 160 | 161 | 160 |
| 0.9 | 144 | 145 | 145 |
| 1.1 | 121 | 118 | 118 |
| 1.3 | 65.5 | 67 | 95 |
| 1.5 | 22 | 45 | 92 |
| 1.7 | 18 | 43 | 91 |
| 2.0 | 16 | 41 | 91 |
| | **Cov. matrix shift** | | |
| Case 1 | 21 | 45 | 93 |
| Case 2 | 20 | 44 | 92 |
| | **Mean and Cov. Matrix shift** | | |
| 0.5 and Case 1 | 146 | 144 | 145 |
| 0.5 and Case 2 | 18 | 43 | 91 |
| 1.5 and Case 1 | 17 | 41 | 91 |
| 1.5 and Case 2 | 16 | 41 | 90 |
| **Mean shift** | | $\boldsymbol{p}$ **= 20000** | |
| 0.9 | 139 | 140 | 140 |
| 1.5 | 19 | 41 | 87 |
| 2.0 | 16 | 39 | 88 |
| | **Cov. matrix shift** | | |
| Case 1 | 19 | 42 | 90 |
| Case 2 | 17 | 42 | 91 |
| | **Mean and Cov. Matrix shift** | | |
| 1.5 and Case 1 | 15 | 40 | 89 |
| 1.5 and Case 2 | 16 | 41 | 88 |

## 4.2. Comparative remarks

We point out the differences with some recently developed high-dimensional change-point monitoring methods. Wu et al. (2022) proposed a data adaptive change-point monitoring method for online sequence of data that requires a stretch of training data without any change-point. We consider the $\mathcal{L}_2$-norm based T1 method from Wu et al. (2022) to compare the robustness under different distributional settings. As in Wu et al. (2022), we consider multivariate normal distributions with $p = 25$, $\rho = 0.2$, 0.8. For skewed distributions, we consider multivariate exponential distributions with IC rate parameter $\lambda$ = 1, without loss of generality, connected by Gaussian and Clayton copula. The size of the training sample is $n = 20$, 50. Critical limits are obtained for a multivariate normal process with $\rho = 0.2$, for training samples of sizes $n = 20$, 50, separately. The nominal $\alpha = 0.1$ is considered as in Wu et al. (2022). We use the same critical limits obtained for multivariate normal with $\rho = 0.2$ for other distributions to obtain the size. The values reported in Table 6 show that the robustness is heavily compromised for Wu et al. (2022).

Since the IC robustness is compromised, an OOC performance comparison with Wu et al. (2022) is not reasonable.

**Table 6: Size of the monitoring method proposed by Wu et al. (2022) under different multivariate settings for dimension $p = 25$ and training sample of size $n = 20$, 50.**

| $p = 25$ | $n = 20$ | $n = 50$ |
|---|---|---|
| | **Multivariate normal** | |
| | Critical limit 0.233 | Critical limit 0.077 |
| $\rho = 0.2$ | 0.107 | 0.104 |
| $\rho = 0.8$ | 0.005 | 0.0003 |
| | **Multivariate exponential** | |
| Gaussian copula | 0.026 | 0.133 |
| Clayton copula $\xi = 5$ | 0.017 | 0.061 |

Holland & Hawkins (2014) proposed a nonparametric online change-point method that does not need a large amount of historical data. However, the method proposed by Holland & Hawkins (2014) requires a larger sample size than the data-dimension. Consequently, this method is not applicable in high-dimensional settings where the data dimension is higher than the sample size. In contrast, the proposed method is applicable to 'small sample, high-dimension' scenario. A robustness comparison is reported in Table 7 between Holland and Hawkins (2014) and the proposed method. QCP indicates the quarantine change-point method by Holland and Hawkins (2014). DFCP indicates the proposed distribution-free change-point method. A similar distributional setting is considered as in Holland and Hawkins (2014). $ARL_0$ values for the QCP method are obtained from Table 2 in Holland and Hawkins (2014). It is observed that the $ARL_0$ values for the QCP method are sometimes far from the nominal choice of 500. The proposed DFCP method provides a more stable IC performance compared to Holland and Hawkins (2014).

**Table 7: $ARL_0$ values for the QCP method and the proposed DFCP method for dimension $p = 5$, 10, under different multivariate settings**.

| | **$p = 5$** | | | | **$p = 10$** | | | |
|---|---|---|---|---|---|---|---|---|
| | **QCP** | | **DFCP** | | **QCP** | | **DFCP** | |
| | **$c = 0$** | **$c = 15$** | **$l_0 = 3$** | **$l_0 = 5$** | **$c = 0$** | **$c = 15$** | **$l_0 = 3$** | **$l_0 = 5$** |
| **Multivariate $t$** | | | | | | | | |
| Cauchy | 54 | 452 | 485.76 | 482.30 | 27 | 394 | 485.43 | 491.99 |
| $t_5$ | 205 | 492 | 487.85 | 497.89 | 92 | 456 | 496.10 | 481.60 |
| **CR multivariate gamma** | | | | | | | | |
| $\rho = 0.9,\ \theta_0 = 4$ | 74 | 455 | 487.62 | 480.35 | 42 | 413 | 485.12 | 478.48 |
| $\rho = 0.9,\ \theta_0 = 2$ | 38 | 439 | 483.73 | 476.83 | 21 | 391 | 494.60 | 495.12 |
| $\rho = 0.9,\ \theta_0 = ½$ | 12 | 410 | 486.44 | 472.86 | 6 | 353 | 493.01 | 481.03 |
| **Transformed multivariate gamma** | | | | | | | | |
| $\rho = 0.9,\ \theta_0 = ½$ | 263 | 479 | 491.55 | 486.12 | 155 | 450 | 489.18 | 483.17 |

### 4.3. Sensitivity analysis

In this section, an empirical sensitivity analysis has been conducted to study the effect of different choices of $l_0$ and $w$ on the change-detection rate. In previous sections, it is observed that different choices of model parameters $l_0$ = 3, 4, 5, and $w$ = 13, 15, 17, do not furnish significant differences in robustness. We consider a multivariate process of dimension $p = 20$ that encountered change point $\tau$ = 50. To assess the detection ability for a symmetric process distribution, we consider a multivariate normal distribution with mean shift $\delta_\mu$ = 1, 1.5, and modified covariance structure (Case I and II) as explained in Section 4.1. For skewed process distribution, we consider a multivariate distribution with endpoints as exponential distributions with rate parameter $\lambda_1$ = 0.001, 5, connected by Clayton copula $\xi$ = 2, 5. A larger range of parameter values and distributional settings could be considered but, for brevity and limited scope of this article, is not included here.

Our objective is to examine the efficacy of the proposed method under different parameter setting in terms of change detection ability. In Table 8, detection rates are reported for different values of $l_0$ and $w$. In standard hypothesis testing problems with 5% significance level, acceptable size-power ratio is 1:18. For the proposed method, from Table 8, this ratio is approximately 1:162, given $\alpha \approx 0.004$. We can conclude that the proposed method has reasonable performance in terms of change detection rate. It is observed that the choices of $w$ do not yield major differences on the detection rate, though $l_0$ = 3 have slightly better detection rate than other $l_0$ values.

**Table 8: Detection rate of the proposed method for different $l_0$ and $w$, for multivariate normal (MVT Norm) distribution and exponential (MVT exponential) distribution.**

**MVT Norm Mean shift**

| | | | |
|---|---|---|---|
| **$\delta_\mu = 1.0$** | | | |
| $w$ | $l_0 = 3$ | $l_0 = 4$ | $l_0 = 5$ |
| 13 | 0.805 | 0.787 | 0.797 |
| 15 | 0.804 | 0.799 | 0.798 |
| 17 | 0.770 | 0.771 | 0.772 |
| **$\delta_\mu = 1.5$** | | | |
| $w$ | $l_0 = 3$ | $l_0 = 4$ | $l_0 = 5$ |
| 13 | 0.733 | 0.711 | 0.713 |
| 15 | 0.709 | 0.706 | 0.686 |
| 17 | 0.677 | 0.681 | 0.665 |
| **MVT Norm Covariance matrix shift** | | | |
| **Case I** | | | |
| $w$ | $l_0 = 3$ | $l_0 = 4$ | $l_0 = 5$ |
| 13 | 0.627 | 0.619 | 0.617 |
| 15 | 0.658 | 0.643 | 0.645 |
| 17 | 0.653 | 0.633 | 0.621 |

| **Case II** | | | |
|---|---|---|---|
| $w$ | $l_0 = 3$ | $l_0 = 4$ | $l_0 = 5$ |
| 13 | 0.619 | 0.623 | 0.621 |
| 15 | 0.647 | 0.638 | 0.632 |
| 17 | 0.642 | 0.648 | 0.639 |
| | **MVT exponential** | | |
| $w$ | $l_0 = 3$ | $l_0 = 4$ | $l_0 = 5$ |
| $\boldsymbol{\xi = 2, \lambda_1 = 0.001}$ | | | |
| 13 | 0.714 | 0.671 | 0.614 |
| 15 | 0.718 | 0.692 | 0.596 |
| 17 | 0.723 | 0.703 | 0.623 |
| $\boldsymbol{\xi = 5, \lambda_1 = 5}$ | | | |
| 13 | 0.729 | 0.669 | 0.602 |
| 15 | 0.722 | 0.686 | 0.593 |
| 17 | 0.750 | 0.711 | 0.626 |

## 5. Illustrative example: Image data monitoring

To illustrate the real-life application of the proposed monitoring procedure, we consider a metal casting dataset (Dabhi et al. 2020). This dataset contains 6633 images of the top view of the submersible pump impeller. Each image is grey-scaled and of size 300×300 pixels. In all images, augmentation is already applied (such as noise reduction etc.). After the vectorization of the images, we obtain a dataset of 6633 random vectors of dimension $p$ = 90000. Clearly, this is a change-point monitoring problem in a *very* high-dimensional setting. Among the 6633 samples, it is known that 2875 samples are IC samples, and 3758 samples are OOC samples. If there is any irregularity in the shape of the product, the part is considered defective (i.e., OOC). We choose random samples of size 50 from both the IC and OOC samples and regard this as a replication of an online production process in which change-point occurs after the 50$^{th}$ sample. In Fig. 2(a) and Fig. 2(b), IC and OOC density plots for the variables 200, 5000 and 10000 are displayed, respectively. It is evident that the variables (mean pixel intensities) do not follow normal distributions, and there is considerable distributional shift present in the dataset after the change-point.

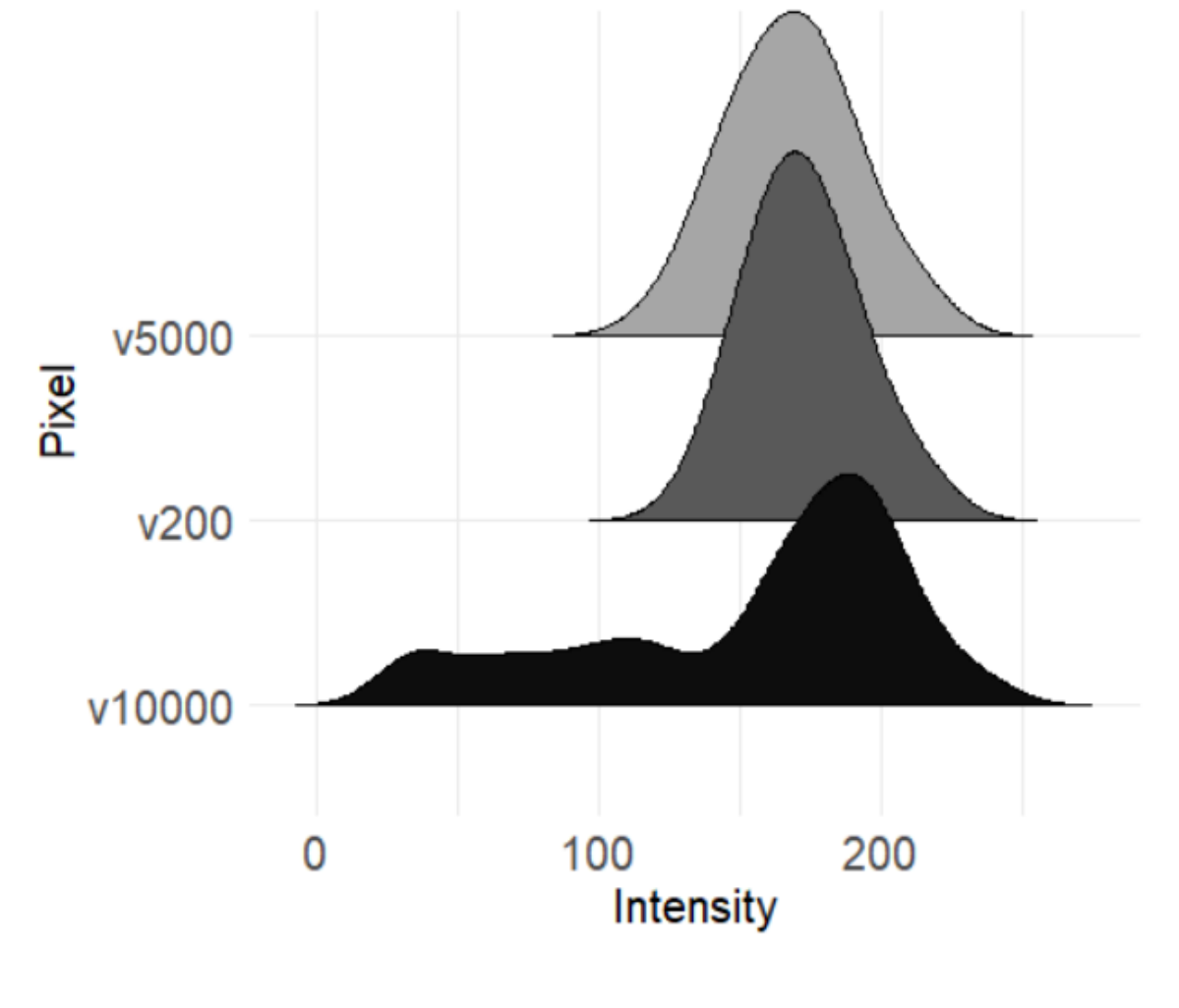

**Fig. 2(a). IC sample**

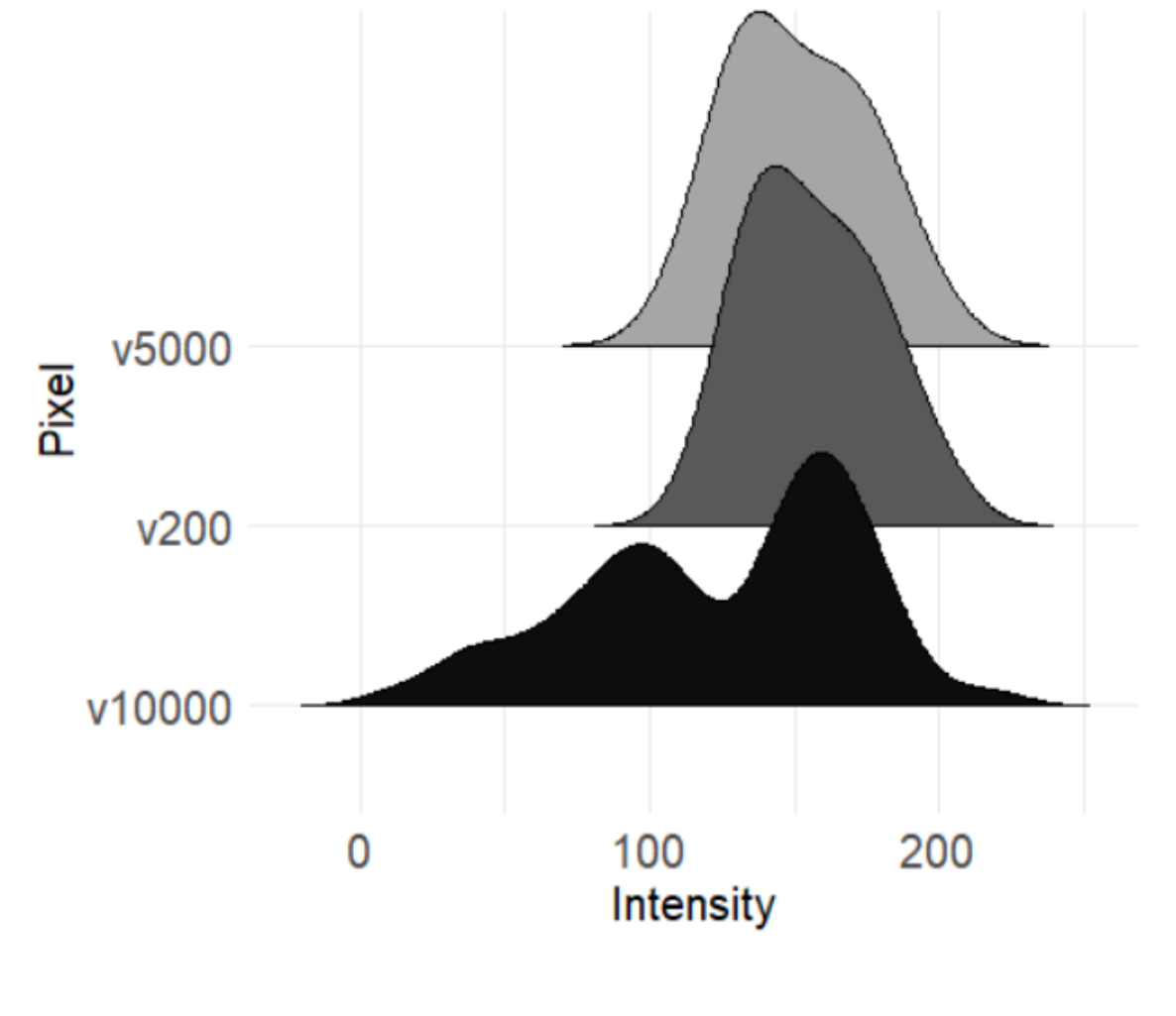

**Fig. 2(b). OOC sample**

**Fig. 2. IC and OOC density plots for the variables 200, 5000 and 10000 from the image data**.

We consider a widow size $w = 15$ and quarantine constant $l_0 = 3$ to monitor the vectorized image sequence. Test statistic values $T_i$ are obtained for every moving window at time instance $i$ = 1, 2, 3, …, and compared with the respective control limits $h_i$ until $T_i \geq h_i$, for some $h_i$. This is plotted in Fig. 3. At $i = 43$, a signal is raised for possible change-point as $T_{43} \geq h_{43}$, i.e., the maximum of $T_{43(0.75)}$ or $(1 - T_{43(0.25)})$ exceeds the control limit. This implies that the 43rd window potentially includes the change-point. For the 43rd window, we have $T_{43(max)}$ = 0.278 and $T_{43(min)} = 0$. Clearly, the minimum of the test statistic values for the 43rd window is more extreme than the maximum. The minimum value occurs at the 5th partition of the 43rd window. Therefore, the estimated change-point is $\hat{\tau} = (43 + 3 + 5 - 1) = 50$. A practitioner should stop the casting process and look for the type of defect when there is a signal at the 43rd window.

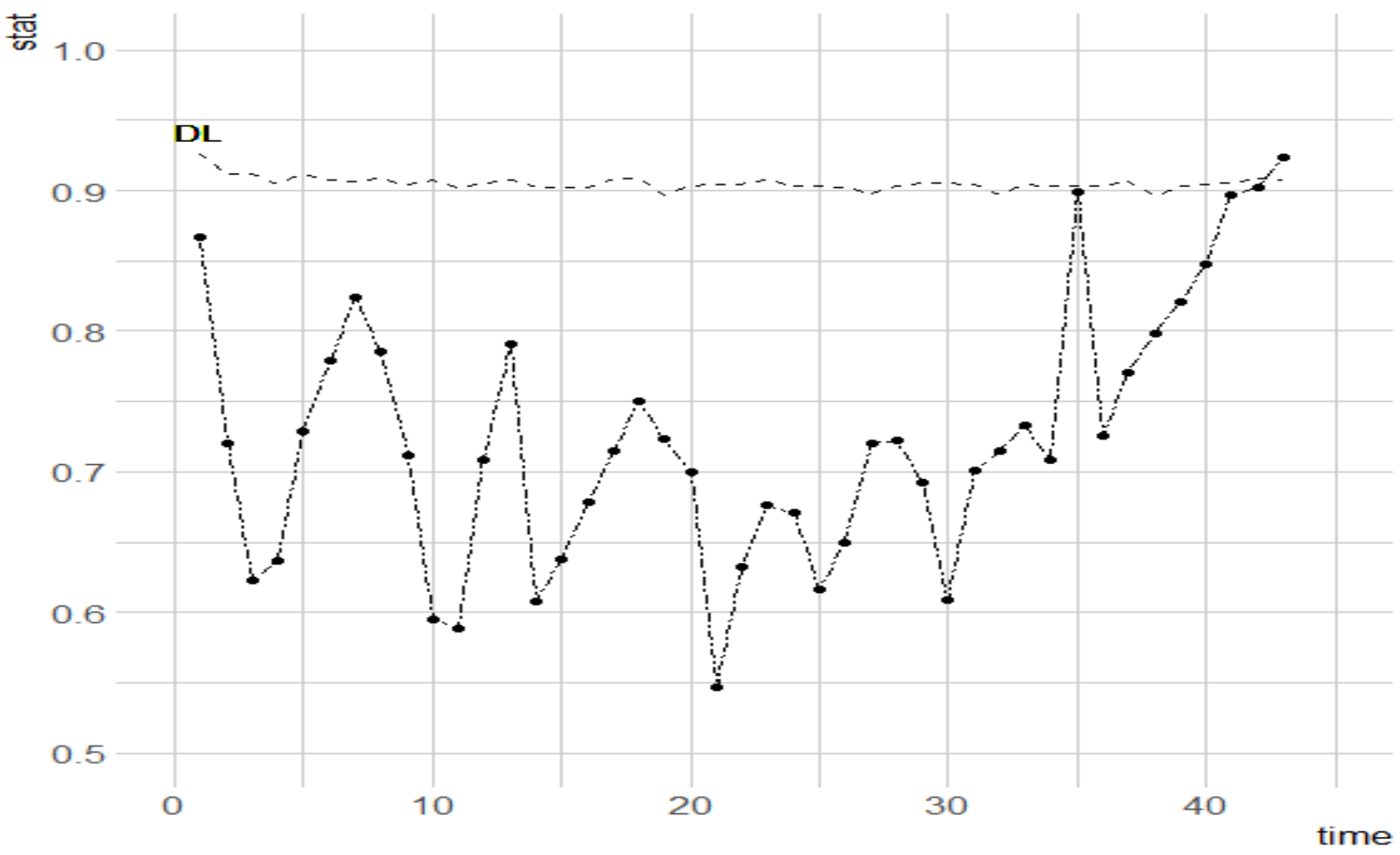

Fig. $3(a)$.

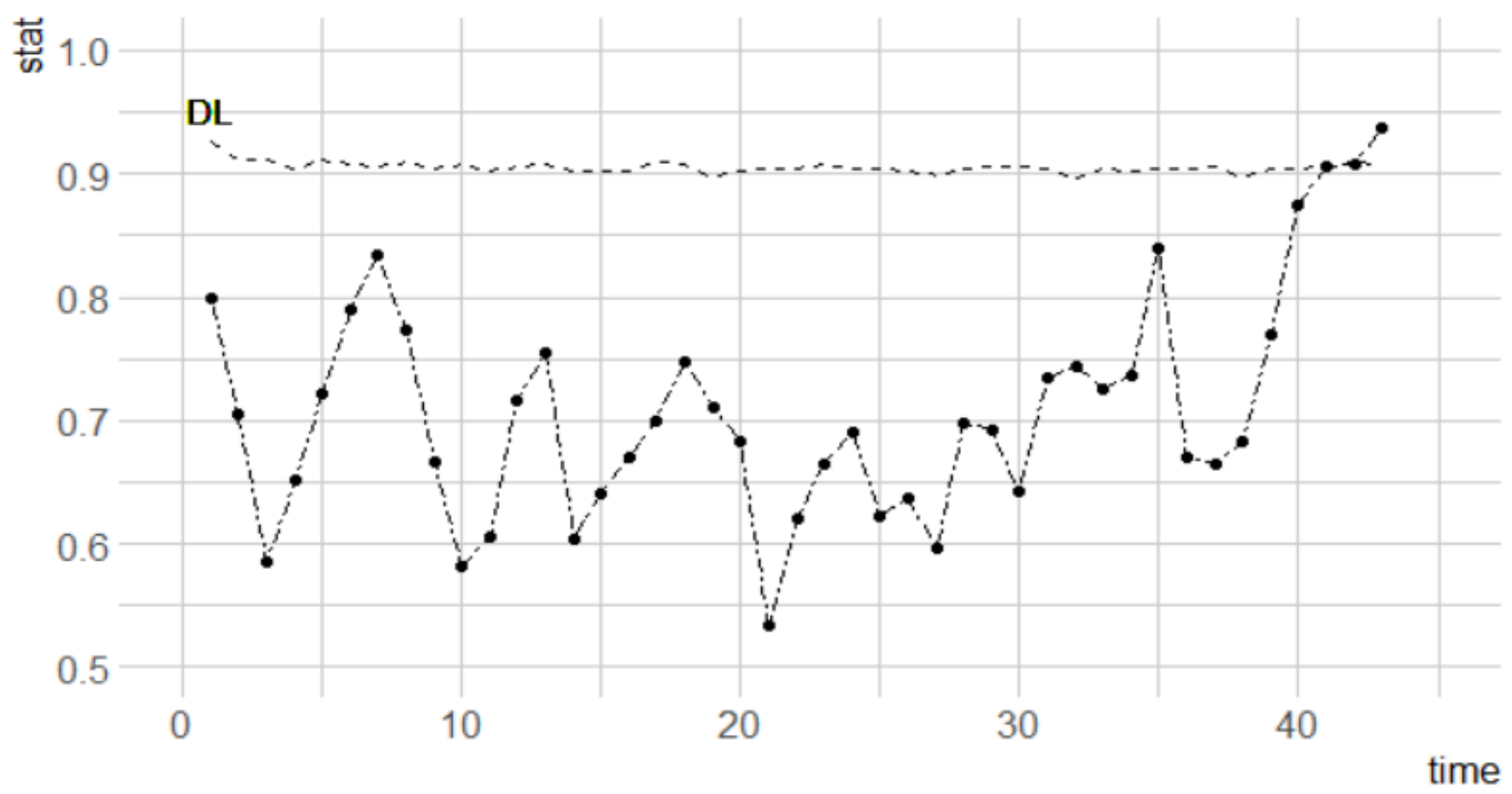

Fig. $3(b)$.

**Fig. 3. Test statistic values for** $(\boldsymbol{a})$ $(l_0, w) = (2,\ 15)$; $(b)$ $(l_0, w) = (3,15)$ **at different time points.**

## 6. Conclusion

In this article, we proposed a distribution-free change-point monitoring procedure applicable in high-dimensional settings. In the early phase and later phase of monitoring, the proposed method is able to detect change-point efficiently for a wide range of shifts, irrespective of the process distribution. With an extensive performance comparison, we conclude that with greater robustness and effective change-detection ability under complex dependence structures, the proposed method should be useful in industrial applications.

## Appendix

### A. Proof of affine invariance.

Let us consider two real valued $p$ -dimensional sequence of random vectors $\boldsymbol{X}$ = $\{\boldsymbol{X_1}, \boldsymbol{X_2}, \boldsymbol{X}, \dots, \boldsymbol{X_m}\}$ and $\boldsymbol{Y}$=$\{\boldsymbol{Y_1}, \boldsymbol{Y_2}, \boldsymbol{Y_3}, \dots, \boldsymbol{Y_n}\}$ obtained at regular time intervals from an online production process/system. Let $\boldsymbol{X}_i^* = \mathcal{T}(\boldsymbol{X}_i) = \boldsymbol{A}\boldsymbol{X}_i+\boldsymbol{b}$ and $\boldsymbol{Y}_j^* = \mathcal{T}(\boldsymbol{Y}_j) = \boldsymbol{A}\boldsymbol{Y}_j+\boldsymbol{b}$ be the affine transformation applied on $\boldsymbol{X}_i$ and $\boldsymbol{Y}_j$ for $i = 1, 2, 3, \dots, m$ and $j = 1, 2, 3, \dots, n$. $\boldsymbol{A}$ is a nonsingular matrix and $\boldsymbol{b}$ is a $p$-dimensional vector. For some $k$ and $s$, suppose $\mathcal{L}_2(\boldsymbol{Y}_k) \leq \mathcal{L}_2(\boldsymbol{Y}_s)$. Since the affine transformation $\mathcal{T}(\boldsymbol{Y}_i) = \boldsymbol{A}\boldsymbol{Y}_i + \boldsymbol{b}$ is a one-to-one transformation, then $\mathcal{L}_2(\mathcal{T}(\boldsymbol{Y}_k)) \leq \mathcal{L}_2(\mathcal{T}(\boldsymbol{Y}_s))$. Similar argument can be provided for $\boldsymbol{X}$ sample. This implies that the relative order will not be affected as long as we take the same affine transformation on $\boldsymbol{X}$ and $\boldsymbol{Y}$. Let us define a $r^{th}$ order precedence statistic as $W_r = \sum_{i=1}^{m} I\left(\mathcal{L}_2(\boldsymbol{X}_i) \leq \mathcal{L}_2\left(\boldsymbol{Y}_{(r)}\right)\right)$, where $I(.)$ is the indicator function. Similarly, we define $W_r^* = \sum_{i=1}^{m} I\left(\mathcal{L}_2(\mathcal{T}(\boldsymbol{X}_i)) \leq \mathcal{L}_2\left(\mathcal{T}\left(\boldsymbol{Y}_{(r)}\right)\right)\right)$. Since the relative order is not affected after taking the affine transformation on $\boldsymbol{X}$ and $\boldsymbol{Y}$, we have $W_r = W_r^*$.

This completes the proof that the proposed method is affine invariant.